\documentclass[final,3p,times,numbers,sort&compress]{elsarticle}

\usepackage{amsthm}  
\newtheorem{theorem}{Theorem}

\usepackage{amsfonts,mathtools,amssymb,siunitx,mathrsfs}
\usepackage{color}
\usepackage[toc,title,page]{appendix}
\usepackage[utf8]{inputenc}
\usepackage{xfrac}
\usepackage{hyperref} 
\hypersetup{colorlinks = true,linkcolor = blue,anchorcolor =red,citecolor = blue,filecolor = red,urlcolor = red,
            pdfauthor=author}
\usepackage{graphicx, nicefrac}

\newcommand{\mi}{\mathrm{i}\,}

\usepackage{braket}

\journal{Physics Letters B}

\begin{document}
\begin{frontmatter}



\title{On the Yamazaki–Hori solution of the Ernst equation}


\author{\color{black}A. R. Melikyan}
\affiliation{organization={Centro Internacional de Física},
            addressline={Instituto de Física, Universidade de Brasília}, 
            city={Brasilia},
            postcode={70910-900}, 
            state={DF},
            country={Brasil}}
\ead{amelik@gmail.com}
\begin{abstract}
A family of solutions to the Ernst equation is presented, which, in a certain limit, recovers the Yamazaki–Hori solution - an extension of the Tomimatsu–Sato solutions for all integer values of the deformation parameter $\delta$. Our solution also recovers, as a special case, the formulation given by Vein, which is equivalent to the Yamazaki–Hori solution. Furthermore, our construction establishes a connection to a nonlinear differential equation proposed by Cosgrove, which is also associated with the Ernst equation.
\end{abstract}

\end{frontmatter}




\section{Introduction}
\label{section:intro}
The Ernst equation  for the complex field $\xi$ (the Ernst potential) has the form:
\begin{align}
    \left(\xi \xi^{*}-1\right) \nabla^{2}\xi=2\xi^{*}\left(\nabla\xi\right)^{2}.\label{intro:Ernst}
\end{align}
It arises in the study of stationary, axially symmetric vacuum solutions to the Einstein field equations and is derived using the Weyl–Lewis–Papapetrou metric \cite{Weyl:1917tb,Lewis:1932ln,Papapetrou:1953db,Islam:2009no}, which describes axisymmetric spacetimes:
\begin{align}
    ds^2 = f(dt - \omega d \varphi)^2 - r^2 f^{-1}
d \varphi - e^{\mu}(dr^2 + dz)^2.\label{intro:WLP_metric}
\end{align}
Here $(r,\varphi,\omega)$  are cylindrical coordinates, and $f=f(r,z),\mu=\mu(r,z),\omega=\omega(r,z)$ 
The relation between the fields is as follows \cite{Islam:2009no}:
\begin{align}
    E=\frac{\xi -1}{\xi +1}; \,\, E=f+\mi  u.\label{intro:Ernst_potential_relations}
\end{align}
The simplest solution for $\xi$ (which corresponds to the $\delta=1$ in the general solution $\xi_{\delta}$ below) reproduces the Kerr solution (see \cite{Islam:2009no} for computational details and \cite{Batic:2023ye} for a recent overview, along with several corrections to earlier works), and has the form:
\begin{align}
    \xi_{1} = px - \mi q y.\label{intro:Kerr}
\end{align}
Tomimatsu  and Sato ($\sf{TS}$) have found other solutions to Ernst equation for $\xi_{\delta}$ are of the type \cite{Tomimatsu:1973sk, Tomimatsu:1972tg}:
\begin{align}
    \xi_{\delta}(x,y;p,q)=\frac{\alpha_{\delta}(x,y;p,q)}{\beta_{\delta}(x,y;p,q)},\label{intro:TS_solutions}
\end{align}
where \( \alpha_{\delta}(x,y; p,q) \) and \( \beta_{\delta}(x,y; p,q) \) are a pair of complex polynomials depending on a deformation  parameter \( \delta \), which controls the degree of multipolar deviation from spherical symmetry, a pair of parameters \( (p, q) \) satisfying the constraint \( p^2 + q^2 = 1 \), and on the prolate spheroidal coordinates \( (x, y) \), which are related to cylindrical coordinates as follows:
\begin{align}
    r=\kappa \left\{(x^2 -1)(1-y^2)\right\}^{1/2}, \quad z= \kappa xy. \label{intro:prolate}
\end{align}
Here, \(\kappa\) is a positive constant related to the mass parameter \(m\), the angular momentum \(J\), and the quadrupole moment \(Q\) as follows:
\begin{align}
\kappa = \frac{m p}{\delta}, \quad J = m^2 q, \quad Q = m^2 \left[q^2 + \frac{(\delta^2 - 1) p^2}{3 \delta^2}\right] \label{intro:kappa_m_J_Q}.
\end{align}
They have discovered non-trivial solutions for $\delta=2,3,4$. For $\delta \geq 2$ the forms of the polynomials become increasingly complex. For example, for $\delta=2$, one has:
\begin{align}
    \xi = \frac{p^2 x^4 + q^2 y^4 - 2 \mi pqxy(x^2-y^2)-1}{2px(x^2-1) - 2 \mi qy(1-y^2)}.\label{intro:TS_n_2}
\end{align}
When \(\delta = 1\), the \(\mathsf{TS}\) solution coincides with the Kerr metric and, in the limit \(p \to 1\), reduces to the Schwarzschild solution. For \(\delta \geq 2\), the \(\mathsf{TS}\) solutions approach static Weyl solutions in the same limit. These represent asymptotically flat, axisymmetric vacuum spacetimes with higher multipole moments, where the deformation parameter \(\delta\) governs the deviation from spherical symmetry \cite{Weyl:1917tb,Islam:2009no}. The \(\mathsf{TS}\) solutions with \(\delta \geq 2\) are generally considered less physically realistic as models of rotating astrophysical objects for several reasons. First, unlike the Kerr solution, their static limit is not spherically symmetric, which is inconsistent with the expected behavior of non-rotating, isolated mass distributions. Second, these spacetimes contain naked singularities (curvature singularities not hidden behind an event horizon), raising physical and causal concerns. In addition, ring singularities and other pathological features such as conical singularities, regions with closed timelike curves, and causality violations can appear \cite{Kodama:2003en,Hoenselaers:1979vq,Gibbons:1973zo}, further complicating their physical interpretation (see \cite{Batic:2023ye} for a recent discussion of the $\delta = 2$ case). Moreover, the multipole moments of the \(\mathsf{TS}\) solutions with \(\delta > 1\) generally do not match those of realistic astrophysical objects, and the physical sources that would give rise to such metrics remain ambiguous \cite{Manko:2012jv,Tanabe:1976gp,Tanabe:1976mc}. Stability analyses also suggest that these solutions may be dynamically unstable under perturbations relevant to astrophysical scenarios. 

Despite these limitations, the \(\mathsf{TS}\) solutions remain of considerable theoretical interest. They are exact, asymptotically flat vacuum solutions of Einstein’s equations, and provide valuable insight into the structure of axisymmetric spacetimes beyond the Kerr family, illustrating the diversity and complexity of the solution space of general relativity. The \(\sf{TS}\) solutions for $\delta=2,3,4$ were later generalized by Yamazaki and Hori \cite{Yamazaki:1977dp,Yamazaki:1977xd,Hori:1978ie}, who constructed solutions for arbitrary integer values of \( \delta \), and Hori further extended the analysis to include arbitrary, not necessarily integer, values of the deformation parameter \( \delta \)~\cite{Hori:1996iz,Hori:1996bb,Hori:1996vt,Hori:1996kp}. In this paper, however, we restrict our attention to integer values of \( \delta=n  \in \mathbb{N} \). 

Taking a different perspective, Cosgrove, whose remarkable analysis includes many interesting results that have yet to be explored in detail and, in particular, allows for the construction of solutions with non-integer values of the deformation parameter $\delta$, studied axially symmetric solutions to the Einstein field equations by analyzing two nonlinear differential equations satisfied by two functions, \( H_4(\eta)\) and \( K(\eta,\nu)\),\footnote{Here, we follow Cosgrove’s original notations.} showing that their properties can be used to reconstruct the Ernst potential \cite{Cosgrove:1978jo,Cosgrove:1978cp,Cosgrove:2008rh, Cosgrove:1977ul, Cosgrove:1981zi,Cosgrove:1982ol, Cosgrove:1977np,Cosgrove:1980wx}.
One of the equations takes the following form:\footnote{The same equation was independently studied by Dale around the same time \cite{Dale:1978xx}, as will be discussed later in the text.}
\begin{align}
      \left[\eta (1+\eta)H^{\prime \prime}_{4}(\eta) \right]^{2} = 4H_{4}^{\prime}(\eta) \left(\eta H_{4}^{\prime}(\eta) -H_{4}(\eta)\right) \left[-\delta^2 +H_{4}(\eta)-(1+\eta)H^{\prime}_{4}(\eta)\right],\label{intro:Cosgrove_H4}
\end{align}
where the coordinates \( (\eta,\nu) \) differ from the prolate spheroidal coordinates \( (x,y) \) defined in \eqref{intro:prolate}, and are instead related to them by the following transformations:
\begin{align}
  \eta={{(x^2 -1)} \big{/}{(1-y^2)}}, \quad \nu=y\big{/}x.\label{intro:Cosgrove_eta_nu}
\end{align}

Our initial motivation was to explore a proof of Nakamura's conjecture \cite{Nakamura:1993ig,Nakamura:1991sv}, which proposes an equivalence between the Ernst equation and the two-dimensional Toda model. Previous attempts to establish this conjecture \cite{Fukuyama:1995cm,Fukuyama:2011ma} have relied on general solutions of the Toda model \cite{Hirota:1988wz,Hirota:1988tt,Ueno:2018gj,Hirota:1988de,Hirota:2009fa} and on verifying the conjecture through Pfaffian identities. In contrast, we adopt the more technically intricate Yamazaki–Hori solution, which offers several advantages over earlier approaches; the details of this will be presented in a forthcoming publication. The present paper emerged as a by-product of that investigation, leading us to construct a family of solutions parametrized by an arbitrary function \( h(x,y,p,q) \), which reduces to the original Yamazaki–Hori solution in a specific limit discussed later. A key advantage of our formulation lies in the flexibility to choose \( h(x,y,p,q) \) in a manner that facilitates the analysis. We further demonstrate that a suitable choice of \( h(x,y,p,q) \) enables a connection with Cosgrove’s differential equation~\eqref{intro:Cosgrove_H4}.

Our paper is organized as follows. In Section \ref{section:hori}, we briefly review the Yamazaki–Hori solution for arbitrary \( n \), along with its equivalent formulation provided by Vein~\cite{Vein:1983if}. In Section \ref{section:general_h_solution}, we present our more general solution, parametrized by an arbitrary function \( h(x,y,p,q) \), and demonstrate that it reduces to the Yamazaki–Hori and Vein solutions as special cases. In Section \ref{section:cosgrove}, we establish a connection to Cosgrove’s differential equation \eqref{intro:Cosgrove_H4} by selecting a specific form of \( h(x,y,p,q) \), which also naturally leads to the unusual coordinate system \( (\eta, \nu) \) \eqref{intro:Cosgrove_eta_nu} employed by Cosgrove. Finally, in the Conclusion, we discuss how our general solution might be used as a basis for approaching a proof of Nakamura’s conjecture.

\section{Yamazaki-Hori solution}
\label{section:hori}
The Yamazaki-Hori solution is quite involved \cite{Yamazaki:1977dp,Yamazaki:1977xd,Hori:1978ie} and will be presented here using the notation of \cite{Vein:1983if,Vein:1980qb}, which will be used throughout this paper. First, the following Hankelian (persymmetric determinant) \cite{Vein:1982jk,Vein:2019nj, Vein:1980qb, Vein:1982ga, Vein:1983ng,Vein:1982xs, Vein:1983if, Vein:2013kv}, denoted by $A=\vert \phi_{m} \vert_{n};\,\,m=0,1, \ldots ,2n-2$, is defined: 
\begin{align}
    A &=\vert a_{ij} \vert_{n}; \, a_{ij} = \phi_{i+j-2},\, i,j=1,2, \ldots ,n.\\
    \phi_{m} &=\frac{1}{m+1}\left[p^{2} (x^2 -1)^{m+1} +q^2 (y^2-1)^{m+1}\right], \,m=0,1, \ldots ,2n-2,  \label{hori:hankelian}
\end{align}
i.e., the Hankel matrix is of the form:
\begin{equation}
A=\left|
\begin{array}{ccccc}
\phi_0 & \phi_1 & \phi_2 & \cdots & \phi_{n-1} \\
\phi_1 & \phi_2 & \phi_3 & \cdots & \phi_{n} \\
\phi_2 & \phi_3 & \phi_4 & \cdots & \phi_{n+1} \\
\vdots & \vdots & \vdots & \ddots & \vdots \\
\phi_{n-1} & \phi_{n} & \phi_{n+1} & \cdots & \phi_{2n-2}
\end{array}
\right|, \quad
\phi_m = \frac{1}{m+1}\left[p^2 (x^2 - 1)^{m+1} + q^2 (y^2 - 1)^{m+1}\right].\label{hori:A_determinant}
\end{equation}
Then, the Yamazaki-Hori solution can be written in the form:
\begin{align}
   \xi_{n}(x,y;p,q)=\frac{\{px\}S_{n}(x,y;p,q\mid x) - \{ \mi qy\}S_{n}(x,y;p,q\mid y)}{S_{n}(x,y;p,q\mid 1)},\label{hori:YH_det_sol}
\end{align}
where $S_{n}(x,y;p,q\mid z)$, is the determinant of the matrix $A$, with the first column replaced by $\left[P_{1}(z),P_{2}(z), \ldots ,P_{n}(z) \right]^{T}$:\footnote{The variable \( z \) in \( S_n(x, y; p, q \mid z) \) specifies the coordinate on which the first column depends.}
\begin{align}
&S_{n}(x,y;p,q\mid z) = 
\left|
\begin{array}{ccccc}
P_1(z) & \phi_1 & \phi_2 & \cdots & \phi_{n-1} \\
P_2(z) & \phi_2 & \phi_3 & \cdots & \phi_{n} \\
P_3(z) & \phi_3 & \phi_4 & \cdots & \phi_{n+1} \\
\vdots & \vdots & \vdots & \ddots & \vdots \\
P_n(z)) & \phi_{n} & \phi_{n+1} & \cdots & \phi_{2n-2}
\end{array}
\right|,\label{hori:YH_solution}
\end{align}
where:
\begin{align}
    P_{k}(z) =\frac{1}{e_{k}}\sum_{l=1}^{k}\, d_{l}\,\left(z^2 -1\right)^{l-1}=\frac{1}{z} \int_0^z \left(t^2 - 1\right)^{k - 1} dt, \label{hori:P_k}
\end{align}
and:
\begin{align}
     e_{l}=(-1)^{l-1}\frac{\Gamma(l+\nicefrac{1}{2})}{\Gamma(\nicefrac{3}{2})(l-1)!}\quad d_{l} &=(-1)^{l-1}\frac{\Gamma(l-\nicefrac{1}{2})}{\Gamma(\nicefrac{1}{2})(l-1)!}, \quad e_{l}=-2ld_{l+1}.
\end{align}
The integral representation of $P_k(z)$ in \eqref{hori:P_k} was obtained in \cite{Vein:1983if}. 

An alternative and significantly simpler form of the Yamazaki–Hori solution was provided by Vein, who demonstrated in \cite{Vein:1983if} that the solution \eqref{hori:YH_det_sol} can be equivalently expressed as follows:
\begin{align}
   \xi_{n}(x,y;p,q)=\frac{\{p\} G_{n}(x,y;p,q\mid x) - \{\mi q\} G_{n}(x,y;p,q\mid y)}{G_{n}(x,y;p,q\mid 1)},\label{hori:Vein_sol}
\end{align}
where:\footnote{As in \( S_{n}(x, y; p, q \mid z) \), the variable \( z \) in \( G_{n}(x, y; p, q \mid z) \) denotes the coordinate on which the last column depends.}
\begin{equation}
G_{n}(x,y;p,q\mid z)=\left|
\begin{array}{cccccc}
\psi_0 & \psi_1 & \cdots & \psi_{n-2} & \psi_{n-1} & \nicefrac{z^1}{1} \\
\psi_1 & \psi_2 & \cdots & \psi_{n-1} & \psi_{n}   & \nicefrac{z^3}{3} \\
\vdots & \vdots & \ddots & \vdots     & \vdots     & \vdots \\ 
\psi_{n-2} & \psi_{n-1} & \cdots & \psi_{2n-4} & \psi_{2n-3} & \nicefrac{z^{2n-3}}{(2n-3)} \\
\psi_{n-1} & \psi_n & \cdots & \psi_{2n-3} & \psi_{2n-2} & \nicefrac{z^{2n-1}}{(2n-1)} \\
1 & 1 & \cdots & 1 & 1 & 0
\end{array}
\right|.\label{hori:Vein_determinants}
\end{equation}
Here the Hankelian matrix \( B = (b_{ij})_{1 \leq i,j \leq n} \) in the main block is formed by the elements:
\begin{align}
    b_{ij} &= \psi_{i+j-2},\, i,j=1,2, \ldots ,n,\\
    \psi_{m} &=\frac{1}{m+1}\left[p^{2} (x^2 )^{m+1} +q^2 (y^2)^{m+1} -1\right], \,m=0,1, \ldots ,2n-2.  \label{hori:Vein_hankelian}
\end{align}
The relation between \(S_{n}(x,y;p,q\mid z)\) in the Yamazaki-Hori's solution \eqref{hori:YH_det_sol} and \(G_{n}(x,y;p,q\mid z)\) in Vein's solution \eqref{hori:Vein_sol} is given in \rm{Theorem 9} of~\cite{Vein:1983if}:
\begin{align}
G_{n}(x,y;p,q\mid z) &=-z\sum_{k,m=1}^{n}\frac{z^{2m-2}}{2m-1}\hat{B}_{km},\label{h:relation_Hori_Vein1}\\
G_{n}(x,y;p,q\mid z) &= -z S_{n}(x,y;p,q\mid z),\label{h:relation_Hori_Vein2}\\
G_{n}(x,y;p,q\mid 1) &= - S_{n}(x,y;p,q\mid 1),\label{h:relation_Hori_Vein3}
\end{align}
The first expression \eqref{h:relation_Hori_Vein1} (for \(z=(x,y)\)) follows from Schur's complement formula (see, for example, \cite{Muir:2023zs,Zhang:2005cb}), and \eqref{h:relation_Hori_Vein2} and \eqref{h:relation_Hori_Vein3} follow from a relation, given in \cite{Vein:2013kv}), between cofactors of the matrices \( B = (b_{ij})_{1 \leq i,j \leq n} \) \eqref{hori:hankelian} and \( B = (b_{ij})_{1 \leq i,j \leq n} \) \eqref{hori:Vein_hankelian}.

Below, we present a more general solution, parametrized by an arbitrary function \( h(x,y;p,q) \), which reduces to Vein's solution and, thus, to the Yamazaki–Hori solution in a specific limit. As a starting point, we state the following theorems concerning persymmetric determinants and sequences that satisfy the generalized Appell equation (the proofs can be found, for example, in \cite{Vein:2013kv}), which will be used throughout the text.
\begin{theorem}
\label{hori:theorem1}
Let $\zeta_{m}$ be a sequences, and $\Delta^{(n)}_{h}$ the $n$-order difference operator with step size $h$, defined as follows:
\begin{align}
    \Delta^{(n)}_{h}\zeta_{0}:=\sum_{k}^{n} \binom{n}{k}\,
    (-h)^{n-k} \zeta_{k}, \quad \zeta_{n}:=\sum_{k}^{n} \binom{n}{k}\,
    (h)^{n-k}\left\{\Delta^{(n)}_{h}\zeta_{0}\right\}.\label{hori:diffrence_op}
\end{align}
Then, the Hankelian formed by $\zeta_{m}$ is invariant under the application of $\Delta^{(n)}_{h}$.
\begin{align}
    \left| \zeta_{m}\right|_{n} =  \left| \Delta^{m}_{h}\zeta_{0}\right|_{n}
\end{align}
\end{theorem}

\begin{theorem}
\label{hori:theorem2}
Let \(\zeta_{m}(x)\) be a sequence that satisfies the generalized Appell equation:
\begin{align}
    \frac{d\zeta_{m}(x)}{dx}=mf(x)\zeta_{m-1}(x),\label{hori:Appel_eq}
\end{align}
and \( \Lambda^{(n)}(x)= \left| \zeta_{m}(x)\right|_{n}\). Then:
\begin{align}
    \frac{d\Lambda^{(n)}(x)}{dx}=\frac{d\zeta_{0}(x)}{dx} \hat{\Lambda}^{(n)}_{11}(x), \label{hori:theorem2_difeq}
\end{align}
where \(\hat{\Lambda}^{(n)}_{11}\) is the cofactor of the element \(\zeta_{11}(x)\)
\end{theorem}

\section{\texorpdfstring{$h$}{h}-dependent solution to the Ernst equation}
\label{section:general_h_solution}

Two key yet straightforward observations serve as the motivation for constructing a more general solution. The first stems from early efforts to prove Nakamura’s conjecture, which posits an equivalence between the Ernst equation and the two-dimensional Toda model. It becomes apparent that the expressions and the proof simplify considerably when \( p = 1, q = 0 \), suggesting that these specific values possess special significance—something that should naturally be captured in a more general formulation. The second, closely related insight arises from a careful review of the main steps in Vein’s derivation of the solution equivalent to the Yamazaki–Hori formulation. This examination reveals that each step of the procedure can be systematically generalized. Both of these observations will be explored in greater detail below.

Let \( C = (c_{ij})_{1 \leq i,j \leq n} \)  be the Hankel matrix formed by the following elements:
\begin{align}
    & c_{ij} = \zeta_{i+j-2}(x,y;p,q),\, i,j=1,2, \ldots ,n,\\
   & \zeta_{m}(x,y;p,q) =\frac{1}{m + 1} \left( p^2 (x^2 - t)^{m + 1} + q^2 (y^2 - t)^{m + 1} - (1 - t)^{m + 1} \right), \,m=0,1, \ldots ,2n-2,  \label{h:zeta_h_hankelian}
\end{align}
where $t$ is  arbitrary for the moment. 
It is easy to check that the application of the difference operator \(\Delta^{(n)}_{h}\) results in the following expression:
\begin{align}
    \Delta^{(n)}_{h}\zeta_{0}(x,y;p,q) = \frac{1}{n + 1} \left[ p^2 (x^2 - t - h)^{n + 1} + q^2 (y^2 - h - h^{\prime})^{n + 1} - (1 - t - h)^{n + 1} + (-h)^{n + 1}(1 - p^2 - q^2). \right]\label{h:dif_op_general}
\end{align}
Due to the constraint \( p^2 +q^2 =1\) the last term vanishes, and we find:
\begin{align}
    \Delta^{(n)}_{h}\zeta_{0}(x,y;p,q) = \frac{1}{n + 1} \left[ p^2 (x^2 - t - h)^{n + 1} + q^2 (y^2 - t - h)^{n + 1} - (1 - t - h)^{n + 1} \right],\label{h:dif_op_general_constraint}
\end{align}
It follows from \eqref{h:dif_op_general_constraint} that its action amounts to a shift of the variable \( t \rightarrow t+h \), and Theorem~\ref{hori:theorem1} implies that the Hankelian formed by the elements \( \zeta_{m}(x,y;p,q) \) remains invariant under the application of the difference operator \( \Delta^{(n)}_{h} \). Comparing \eqref{hori:hankelian}, \eqref{hori:Vein_hankelian} with \eqref{h:dif_op_general}, we observe that the Hankel matrices in the Yamazaki-Hori and Vein's solutions correspond to \( t = 1 \) and \( t = 0 \), respectively. We will not fix this parameter, but instead derive the necessary expression in this general setting first. In the next section, we will choose the function \( h \) in such a way as to establish a connection with Cosgrove's differential equation \eqref{intro:Cosgrove_H4}.

To proceed, we need several expressions, relating a sequence \(\theta_{k} \) with the sequence \( \zeta_{m}\) obtained as a result of the application $m$-order difference operator with step size $h$ \eqref{hori:diffrence_op}:
\begin{align}
   \zeta_{m}:=\sum_{k}^{m} \binom{m}{k}\,
    (-h)^{m-k} \theta_{k} \label{h:relation_zeta_theta}
\end{align}
Then, one can show the following identities:\footnote{These formulas match those given in \cite{Vein:1983ng} for the particular case \(h=1\)}
\begin{align}
   (h)^{-n}  \sum_{r=0}^{m} \binom{m}{r}\,(h)^{-r}\zeta_{r+n} &=(h)^{-m} \sum_{r=0}^{n} (-1)^{n+r} \binom{n}{r}\,(h)^{-r}\theta_{r+m},\label{h:binom_identity1}\\
    \sum_{r=0}^{m} (-1)^{r} (h)^{-r}\zeta_{r} &=\sum_{s=0}^{m}\,(-1)^{s} \binom{m+1}{s+1}\,(h)^{-s}\theta_{s},\label{h:binom_identity2}\\
  \sum_{r=0}^{m} (h)^{-r}\theta_{r} &=\sum_{s=0}^{m} \binom{m+1}{s+1}\,(h)^{-s}\zeta_{s}\label{h:binom_identity3},
\end{align}
Next, we define the following generalized Pascal matrix, which has many interesting combinatorial and algebraic properties \cite{Call:1993vh,Zhang:1998ub,Zhang:1997ah,Stanimirovic:2011py}:
\begin{align}
\left[\mathcal{P}^{(h)}_n\right]_{ij} &=
\begin{cases}
h^{i-j} \binom{i-1}{j-1}, & \text{if } i \geq j \\
0, & \text{if } i < j
\end{cases} \quad \text{where } i,j = 1,2,\ldots,n,\label{h:Pascal_matrix}\\
\mathcal{P}^{(h)}_n &=
\begin{pmatrix}
1 & 0 & 0 & \cdots & 0 \\
h & 1 & 0 & \cdots & 0 \\
h^2 & 2h & 1 & \cdots & 0 \\
h^3 & 3h^2 & 3h & 1 & \cdots \\
\vdots & \vdots & \vdots & \vdots & \ddots \\
h^{n-2} & (n\!-\!2)h^{n-3} & \cdots & \binom{n-2}{1}h & 1 & 0 \\
h^{n-1} & (n\!-\!1)h^{n-2} & \cdots & \binom{n-1}{2}h^2 & \binom{n-1}{1}h & 1
\end{pmatrix}.
\end{align}
The inverse matrix can be written in the form:
\[
\left[(\mathcal{P}^{(h)}_n)^{-1}\right]_{ij} =
\begin{cases}
(-1)^{i-j} h^{i-j} \binom{i-1}{j-1}, & \text{if } i \geq j \\
0, & \text{if } i < j
\end{cases}\label{h:Pascal_matrix_inverse}
\]
It is then easy to verify, using the identities \eqref{h:binom_identity1} - \eqref{h:binom_identity3} and the form of the generalized Pascal matrix \eqref{h:Pascal_matrix}, that the matrices \( B = (b_{ij})_{1 \leq i,j \leq n};\, b_{ij} = \theta_{i+j-2},\, i,j = 1, 2, \ldots ,n \) and \( C = (c_{ij})_{1 \leq i,j \leq n};\, c_{ij} = \zeta_{i+j-2},\, i,j = 1, 2, \ldots ,n \) satisfy the relation:
\begin{align}
    \mathcal{P}^{(h)}_n C (\mathcal{P}^{(h)}_n)^{T} =B.\label{h:PCPT_is_B}
\end{align}
Written explicitly, this is equivalent to:
\begin{align}
b_{ij} &=\sum_{k,l=1}^{n}\, (h)^{i+j-k-l}\,\binom{i-1}{k-1}\binom{j-1}{l-1}\, c_{kl} \label{h_b_intermsof_c}
\\
c_{ij} &=\sum_{k,l=1}^{n}\, (-1)^{i+j+k+l}\,(h)^{i+j-k-l}\,\binom{i-1}{k-1}\binom{j-1}{l-1}\, b_{kl}. \label{h:h_c_intermsof_b}
\end{align}
Finally, the preceding formulas also allow us to establish a relationship between the cofactors \( \hat{B} \) and \( \hat{C} \) of the matrices \( B \) and \( C \):\footnote{We also use here the relation $\rm{det}(B)=\rm{det}(C)$, which follows from \eqref{h:h_c_intermsof_b}}
\begin{align}
\hat{B}_{ij} &=\sum_{k,\ell=1}^{n}\,(-1)^{i+j+k+\ell}\,(h)^{i+j-k-\ell}\,\binom{i-1}{k-1}\binom{j-1}{\ell-1}\, \hat{C}_{k\ell} \label{h:h_cof_b_intermsof_cof_c}
\\
\hat{C}_{ij} &=\sum_{k,\ell=1}^{n}\,(h)^{i+j-k-\ell}\,\binom{k-1}{i-1}\binom{\ell-1}{j-1}\, \hat{B}_{k \ell}. \label{h:h_cof_c_intermsof_cof_b}
\end{align}

We are now ready to present our general \( h \)-dependent solution of the Ernst equation, where \( h = h(x,y;p,q) \) is an arbitrary function. First, as mentioned above, the formulas \eqref{hori:hankelian}, \eqref{hori:Vein_hankelian}, and \eqref{h:zeta_h_hankelian} imply that the Yamazaki-Hori solution corresponds to \( t = 1 \), while Vein's solution corresponds to \( t = 0 \). The relation between the determinants in \eqref{hori:YH_det_sol} and \eqref{hori:Vein_sol} is given by the expressions \eqref{h:relation_Hori_Vein1} - \eqref{h:relation_Hori_Vein3}.
Let us begin with Vein's solution, i.e., by setting \( t = 0 \) in \eqref{h:zeta_h_hankelian}. Thus, from \eqref{hori:Vein_hankelian}, we obtain the matrix \( B = (b_{ij})_{1 \leq i,j \leq n} \), formed by the elements:
\begin{align}
    b_{ij} &= \psi_{i+j-2},\, i,j=1,2, \ldots ,n,\\
    \psi_{m} &=\frac{1}{m+1}\left[p^{2} (x^2 )^{m+1} +q^2 (y^2)^{m+1} -1\right], \,m=0,1, \ldots ,2n-2,  \label{h:Vein_hankelian}
\end{align}
We now apply the difference operator \(\Delta^{(n)}_{h}\) \eqref{h:dif_op_general}. The result follows from \eqref{h:dif_op_general_constraint} with $t=0$:
\begin{align}
    \zeta_{n}(h)=\Delta^{(n)}_{h}\psi_{0}(x,y;p,q) = \frac{1}{n + 1} \left[ p^2 (x^2 - h)^{n + 1} + q^2 (y^2 - h)^{n + 1} - (1 - h)^{n + 1} \right].\label{h:dif_op_Vein_psi}
\end{align}
This effectively leads us back to the general expression \eqref{h:zeta_h_hankelian}. We thus conclude that the relation between Vein's \( \psi_{m} \) in \eqref{h:Vein_hankelian} and the general \(\zeta_{n}  \) in \eqref{h:dif_op_Vein_psi}  is established by \eqref{h:relation_zeta_theta}, which is the only condition needed to derive the identities \eqref{h:binom_identity1} through \eqref{h:h_cof_c_intermsof_cof_b}. By recalling \eqref{h:relation_Hori_Vein1} and making use of \eqref{h:h_cof_b_intermsof_cof_c}, we arrive at:
\begin{align}
    & S_{n}(x,y;p,q\mid z) =\sum_{k,\ell=1}^{n}\frac{z^{2\ell-2}}{2\ell-2}\hat{B}_{k\ell} \\
    &= \sum_{k,\ell=1}^{n}\,\hat{C}_{k \ell}\,
    \left\{ \sum_{r=1}^{n}\,(-1)^{k+r}\,(h)^{k-r}\,\binom{k-1}{r-1}\right\} \left\{ \sum_{s=1}^{n}\frac{z^{2s-2}}{2s-1}\,(-1)^{\ell-s}\,(h)^{\ell-s}\binom{\ell-1}{s-1}\right\}.
\end{align}
The expression inside the first pair of curly brackets corresponds to the binomial expansion of \( (1 - h)^{k - 1} \), while the expression inside the second can be identified as Gauss’s hypergeometric function: 
\( (-h)^{\ell - 1} \cdot {}_2F_1\left( \frac{1}{2}, 1 - \ell; \frac{3}{2}; \frac{z^2}{h} \right) \). Hence, we find:
\begin{align}
    & S_{n}(x,y;p,q\mid z)=:S^{(h)}_{n}(x,y;p,q\mid z)= \sum_{k,l=1}^{n}\,\hat{C}_{k \ell}\,
    \left\{ (1-h)^{k-1} \right\} \left\{ {(-h)^{\ell - 1}} \cdot {}_2F_1\left( \frac{1}{2}, 1 - \ell; \frac{3}{2}; \frac{z^2}{h} \right).\label{h:bordered_hypergeom}
 \right\}.
\end{align}
Once again making use of the bordered determinant (Schur’s complement) formula, and denoting:\footnote{We can also write the result in terms of Jacobi polynomials $P_{n}^{\left(\alpha,\, \beta\right)}$:
\[
(-h)^{\ell - 1} \, {}_2F_1\left( \frac{1}{2}, 1 - \ell; \frac{3}{2}; \frac{z^2}{h} \right) 
= (-h)^{\ell - 1} \cdot \frac{(\ell - 1)!}{\left(\frac{3}{2}\right)_{\ell - 1}} \, P_{\ell - 1}^{\left(\frac{1}{2},\, -\ell\right)}\left(1 - \frac{2z^2}{h}\right)\label{h:Jacobi}
\]
}
\begin{align}
    \hat{F}(\ell,h;z):={(-h)^{\ell - 1}} \cdot {}_2F_1\left( \frac{1}{2}, 1 - \ell; \frac{3}{2}; \frac{z^2}{h} \right),\label{h:F_hat}
\end{align}
we arrive at the following result from the equation above:
\begin{equation}
S^{(h)}_{n}(x,y;p,q\mid z)=\left|
\begin{array}{cccccc}
\zeta_0(h) & \zeta_1(h) & \cdots & \zeta_{n-2}(h) & \zeta_{n-1}(h) & \hat{F}(1,h;z)  \\
\zeta_1(h) & \zeta_2(h) & \cdots & \zeta_{n-1}(h) & \zeta_{n}(h)   & \hat{F}(2,h;z) \\
\vdots & \vdots & \ddots & \vdots     & \vdots     & \vdots \\ 
\zeta_{n-2}(h) & \zeta_{n-1}(h) & \cdots & \zeta_{2n-4}(h) & \zeta_{2n-3}(h) & \hat{F}(n-2,h;z) \\
\zeta_{n-1}(h) & \zeta_n(h) & \cdots & \zeta_{2n-3}(h) & \zeta_{2n-2}(h) & \hat{F}(n-1,h;z) \\
(1-h)^{0} & (1-h)^{1} & \cdots & (1-h)^{n-2} & (1-h)^{n-1} & 0
\end{array}
\right|.\label{h:hypergeom_determinants}
\end{equation}
Thus, we obtain the solution of the form:
\begin{align}
   \xi_{n}(x,y;p,q)=\frac{\{p x\} S^{(h)}_{n}(x,y;p,q\mid x) - \{\mi q y\} S^{(h)}_{n}(x,y;p,q\mid y)}{S^{(h)}_{n}(x,y;p,q\mid 1)},\label{h:my_h_sol}
\end{align}
To relate our general solution to the one obtained by Vein (which is equivalent to the original Yamazaki-Hori solution), we note that $\zeta_{n}(h)$ in equation \eqref{h:dif_op_Vein_psi} coincides with $\psi_{n}$ from \eqref{h:Vein_hankelian} in the limit $h \rightarrow 0$.
It is also easy to check that in this limit, we have:
\[
\lim_{h \to 0}\left\{ {(-h)^{\ell - 1}} \cdot {}_2F_1\left( \frac{1}{2}, 1 - \ell; \frac{3}{2}; \frac{z^2}{h} \right)\right\} = \frac{z^{2(\ell - 1)}}{2\ell - 1}.
\]
Thus, we conclude that our general 
$h$-dependent solution for the determinant \eqref{h:hypergeom_determinants} precisely matches the expression in \eqref{hori:Vein_determinants}, thus reproducing the result obtained by Vein.

We can also directly obtain the Yamazaki-Hori solution, by considering the limit $h \rightarrow 1$. In this case, $\zeta_{n}(h)$ in  \eqref{h:dif_op_Vein_psi} coincides with $\phi_{n}$ \eqref{hori:hankelian}, and it is readily checked that:
\[
\lim_{h \to 1}\left\{(-h)^{\ell-1} \, {}_2F_1\left( \frac{1}{2}, 1-\ell; \frac{3}{2}; \frac{z^2}{h} \right)\right\}
= \frac{1}{z} \int_0^z \left(t^2 - 1\right)^{\ell - 1} dt,\label{h:recover_YH_P_k}
\]
which is precisely the functions $P_l(z)$ \eqref{hori:P_k} from the original Yamazaki-Hori solution. Thus, using the notation \eqref{h:F_hat} for $\hat{F}(\ell,h;z)$ we arrive at the relation:
\begin{align}
\hat{F}(\ell,1;z)=P_{\ell}(z). \label{F_hat_is_Pk}
\end{align}
As $h \rightarrow 1$, all terms in the last row of \eqref{h:hypergeom_determinants}, except for the first element, vanish. Consequently, this leads to the immediate recovery of the Yamazaki-Hori solution \eqref{hori:YH_det_sol}.

To summarize, we have constructed a solution that recovers both Vein’s and Yamazaki–Hori solutions in the limits \( h \to 0 \) and \( h \to 1 \), respectively. The key advantage of our approach is the flexibility in choosing the function \( h(x, y) \), although particular applications may require imposing constraints on it. This situation closely resembles the two-dimensional Toda model. Recall that the two-dimensional Toda model, written in terms of the tau-function $\tau_{n}$ \cite{Harnad:2021av,Hirota:2009fa}, has the form:
\begin{align}
    \frac{\partial^{2} \tau_{n}}{\partial x \, \partial y}\cdot\tau_{n} - \frac{\partial \tau_{n} \, \partial \tau_{n}}{\partial x \, \partial y}=\tau_{n-1}\cdot\tau_{n+1}.\label{hori:Toda}
\end{align}
The general solution is given by the following expression \cite{Leznov:1981bf,Hirota:2009fa}:
\begin{align}
    \tau_{n}=\rm{Det} \left[\left(\frac{\partial}{\partial x}\right)^{i-1} \left(\frac{\partial y}{\partial y}\right)^{j-1}\,\psi(x,y)\right]\;,\quad{1 \leq i,j \leq n},\label{hori:Toda_psi_sol}
\end{align}
where $\psi(x,y)$ is an arbitrary function. Although the general solution depends on an arbitrary function \( \psi(x, y) \), compatibility with Nakamura’s conjecture requires restricting \( \psi \) to a specific variant known as the two-dimensional Toda molecule \cite{Hirota:1988wz}. This variant corresponds to boundary conditions that limit \( \psi(x, y) \) to a finite sum of products of functions, each depending on a single variable. Moreover, existing verifications of special cases further specify the form of \( \psi \) \cite{Fukuyama:1995cm,Fukuyama:2011ma}, suggesting that a similar restriction on \( h(x, y) \) may be necessary to address the conjecture for all values of the parameter \( p \).

A second, and more significant, observation is that no independent derivation of the Yamazaki–Hori solutions currently exists. In the next section, we investigate a particular choice of \( h(x, y) \) that leads to Cosgrove’s equation, which plays a central role in reconstructing the Ernst potential. It is worth noting that Cosgrove’s equation is a special case of the \(\sigma \)-form of the Painlevé $\rm{VI}$ equation, for which a large class of rational solutions is known~\cite{Okamoto:1986cf}. By choosing \( h(x, y) \) appropriately, as we do in the next section, it should be possible to connect these rational Painlevé $\rm{VI}$ solutions to the Hankelian framework described in Theorem~\ref{cosgrove:Dale_theorem1}, providing a direct pathway to derive the Yamazaki–Hori solutions from first principles.

\section{Connection to Cosgrove’s differential equation}
\label{section:cosgrove}

Various choices for the function $h$ in \eqref{h:hypergeom_determinants} can be explored. The appearance of the hypergeometric function does not, in itself, introduce additional complexity, as is evident from the bordered determinant formula \eqref{h:bordered_hypergeom}. It is also worth noting that determinants involving hypergeometric functions have previously been considered, for example, by Burchnall in \cite{Burchnall:1954ug}. The function $h$ should instead be chosen to either yield a closed-form expression or to significantly simplify the cofactors $\hat{C}_{k \ell}$, thereby facilitating explicit computations. This consideration is particularly crucial for verifying, for instance, Nakamura's conjecture, where the core of the problem effectively reduces to checking certain differential equations.\footnote{This will be the subject of our forthcoming publication.}. With this in mind, we start by considering our general solution \eqref{h:bordered_hypergeom} for $z=x$, and choose $h=y^2$. Then, the last column of \eqref{h:bordered_hypergeom} is formed by $\hat{F}(\ell,1;x/y)=P_{\ell}(x/y);\ell = 1,2,\ldots ,n$, and the persymmetric determinant of the main matrix in \eqref{h:bordered_hypergeom}, \({\mathcal{H}}_{n}=\vert {\hat{\zeta}}_m(x,y;p) \vert_{n};\quad m = 0, 1, \dots, 2n - 2\), takes the form:
\begin{align}
\mathcal{H}_n = 
\left|
\begin{array}{cccccc}
\hat{\zeta}_0     & \hat{\zeta}_1     & \hat{\zeta}_2     & \cdots & \zeta_{n-1} \\
\hat{\zeta}_1     & \hat{\zeta}_2     & \hat{\zeta}_3     & \cdots & \hat{\zeta}_n \\
\hat{\zeta}_2     & \hat{\zeta}_3     & \hat{\zeta}_4     & \cdots & \hat{\zeta}_{n+1} \\
\vdots      & \vdots      & \vdots      & \ddots & \vdots \\
\hat{\zeta}_{n-1} & \hat{\zeta}_n     & \hat{\zeta}_{n+1} & \cdots & \hat{\zeta}_{2n-2},
\end{array}
\right|\label{cosgrove:main_matrix_hankelian}
\end{align}
where 
\begin{align}
\hat{\zeta}_m(x,y;p):=\zeta_m(y^2) = \frac{1}{m + 1} \left( p^2 (x^2 - y^2)^{m + 1}  - (1 - y^2)^{m + 1} \right), \quad m = 0, 1, \dots, 2n - 2.\label{cosgrove:hat_zeta}
\end{align}
The main motivation behind this choice of the function \( h \) stems from the proof of Nakamura's conjecture in the case where \( (p = 1, q = 0) \) \cite{Fukuyama:1995cm,Fukuyama:2011ma}. As previously noted, the expressions simplify considerably in this case. It is therefore advantageous to work with formulas that initially involve only the parameter \( p \). The choice \( h = y^2 \) serves this purpose, as we will demonstrate below, allowing the general formulas to be expressed solely in terms of \( p \) from the outset. To this end, we observe that the function:

\begin{align}
    g_{m}(x,y;h):=\frac{\hat{\zeta}_m(x,y;p)}{(x^2 -1)^{(m+1)}}
\end{align}
satisfies the Appell equations \eqref{hori:Appel_eq}:
\begin{align}
   \frac{\partial g_{m}(x,y;h)}{\partial x} &=mF_{1}g_{m-1}(x,y;h);\quad  F_{1}=-\frac{2  x (1 - y^2)}{(x^2-1)^2}\label{cosrgove:Appell_F1}\\
   \frac{\partial g_{m}(x,y;h)}{\partial y} &=mF_{2}g_{m-1}(x,y;h); \quad  F_{2}=-\frac{2 y}{x^2-1}\label{cosrgove:Appell_F2}
\end{align}
Hence, Theorem~\ref{hori:theorem2} applies, leading to the following equations:
\begin{align}
\left(x^2 -1 \right)\frac{\partial \mathcal{H}_n}{\partial x} &=2x n^{2} \mathcal{H}_n +2x\left(1-p^{2}\right) \left(1-y^{2}\right)\left(\mathcal{H}_{n}\right)_{11},\label{cosgrove:dif_eq_Hn_x}\\
\frac{\partial \mathcal{H}_n}{\partial y} &=2y\left(1-p^{2}\right)\left(\mathcal{H}_{n}\right)_{11},\label{cosgrove:dif_eq_Hn_y}
\end{align}
These equations explicitly illustrate that the case \( p = 1 \) is special and significantly simpler. The general solution is given by:
\begin{align}
    \mathcal{H}_n(x, y, p) = (x^2 - 1)^{n^2} R\left( \frac{1 - y^2}{x^2 - 1}, p \right),\label{cosgrove:Solution_PDE} 
\end{align}
where $R$ is an arbitrary function. Determining the exact form of $R$ remains an open problem. However, observing its dependence on Cosgrove’s variable  \( \eta={(x^2 - 1)}\big{/}{(1 - y^2)}\) (see equation \eqref{intro:Cosgrove_eta_nu}),  we can gain further insight by applying elementary row and column operations to the matrix in \eqref{cosgrove:main_matrix_hankelian}, which leads to:
\begin{equation}
    \mathcal{H}_{n}=\left(1-y^2 \right)^{n^{2}}\left(p\right)^{n(n+1)} \hat{\mathcal{H}}_{n},
\end{equation}
where $\hat{\mathcal{H}}_{n}$ is a persymmetric determinant $\hat{\mathcal{H}}_{n}=\hat{\mathcal{H}}_{n}=\vert {\xi}_m(x,y;p) \vert_{n};\quad m = 0, 1, \dots, 2n - 2.$ 
with the elements: 
\begin{align}
&\hat{\mathcal{H}}_{n}=\vert {\xi}_m(x,y;p) \vert_{n},\\
&{\xi}_m(x,y;p): = \frac{1}{m + 1} \left\{ \left(\frac{x^2 - y^2}{1 - y^2}\right)^{m + 1}   - {1}\big{/}{p^2} \right\}, \quad m = 0, 1, \dots, 2n - 2.\label{cosgrove:xi}
\end{align}
By applying Theorem~\ref{hori:theorem1} and acting with the $n$-th order difference operator of step size $h'$ on ${\xi}_m(x,y;p)$, we obtain:
\[
D^{(n)}_{h^{\prime}}[{\xi}_0](x,y;p) = \frac{1}{n + 1} \left[ \left(\frac{x^2 - y^2}{1 - y^2} - h^{\prime}\right)^{n + 1} - \frac{(1 - h^{\prime})^{n + 1}}{p^2} - (-h^{\prime})^{n + 1} \left(1 - \frac{1}{p^2}\right) \right].
\]
We consider the following two cases of interest, whose relevance to our analysis will become clear shortly:
\begin{align}
&\text{(i) } h^{\prime}_{1}=1: \quad \quad \quad\,\,\, D^{(n)}_{1}[{\xi}_0](\eta;p) =   \frac{1}{n + 1} \left[ \eta ^{n + 1} + (-1)^{n} \left(1-{1}\big{/}{p^2} \right) \right],\label{cosgrove:chi_case_1}\\
&\text{(ii) } h^{\prime}_{2}=\frac{x^2 - y^2}{1 - y^2}: \,\,\,\,D^{(n)}_{2}[{\xi}_0](\eta;p) =\frac{(-1)^n(1-1/p^2)}{(n+1)} \left[ (\eta + 1)^{n+1} - \left( 1-p^2\right)^{-1}\eta^{n+1}\right].
\label{cosgrove:chi_case_2}
\end{align}
where we have denoted: \( \eta={(x^2 - 1)}\big{/}{(1 - y^2)}\). Note, that this is consistent with the coordinate $\eta$ employed by Cosgrove to investigate the differential equation \eqref{intro:Cosgrove_H4}. In addition, as noted earlier, the last column of our general solution \eqref{h:bordered_hypergeom}, when fixing $h=y^2$, is formed by $\hat{F}(\ell,1;x/y)=P_{\ell}(x/y)=P_{\ell}(1/\nu);\ell = 1,2,\ldots ,n$. Thus, our general solution (ignoring an overall factor multiplying the determinant) reduces to a determinant which depends only on Cosgrove's coordinates $(\eta,\nu)$ (cf. \eqref{intro:Cosgrove_eta_nu}).

We will now state two theorems: the first was established by Dale in \cite{Dale:1978xx}, and the second by Dale and Vein in \cite{Vein:1980qb}:
\begin{theorem}
\label{cosgrove:Dale_theorem1}
A solution to the Cosgrove-Dale equation \eqref{intro:Cosgrove_H4} for $H_{4}(\eta)$:
\begin{align}
    \left[\eta (1+\eta)H^{\prime \prime}_{4}(\eta) \right]^{2} = 4H_{4}^{\prime}(\eta) \left(\eta H_{4}^{\prime}(\eta) -H_{4}(\eta)\right) \left[-\delta^2 +H_{4}(\eta)-(1+\eta)H^{\prime}_{4}(\eta)\right],\label{cosgrove:Cosgrove_H4}
\end{align}
with $\delta=n;n=1,2,\ldots,$  is given by the following function:
\begin{align}
    H_{4}(\eta) = (1-c) \eta \hat{D}_{11}\big{/}det{(D)},
\end{align}
where $c$ is an arbitrary constant, $D$ stands for the Hankelian $D=\vert \alpha_{m} \vert_{n}$, and $\hat{D}_{11}$ is the corresponding cofactor:
\begin{align}
    \alpha_m(\eta):=\frac{1}{m+1}\left[ \eta ^{m + 1} + (-1)^{m} c \right].
\end{align}
\end{theorem}
A few remarks are in order: First, upon comparing the Hankelian constructed from $\alpha_m(\eta)$, as presented in the theorem above, with the Hankelian in \eqref{cosgrove:chi_case_1} derived from our general solution \eqref{h:bordered_hypergeom}, we find that the two are identical, provided the following identification:
\begin{align}
c=\left(1-{1}\big{/}{p^2}\right)\label{cosgrove:c_in_p}
\end{align}
The second remark is that the theorem applies for $\delta = n$, where $n = 1, 2, \ldots$, which precisely corresponds to the Tomimatsu--Sato solutions for $\delta = 1, 2, 3, 4$, and was later generalized by Yamazaki and Hori to arbitrary positive integer values of $\delta$. We now turn to the second relevant theorem.
\begin{theorem}
\label{cosgrove:Dale_theorem2}
The Hankelian: 
\begin{align}
    \beta_m(\eta):=\frac{1}{m+1}\left[ \left(1+\eta\right) ^{m + 1} - (\eta)^{m + 1}  \right]
\end{align}
satisfies the following differential equation:
\begin{align}
\left\{\left[\eta (1+\eta)f(\eta) \right]^{\prime \prime}\right\}^{2} = 4n^2 \left\{\eta f(\eta)\right\}^{\prime}  \left\{\left(1+ \eta \right) f(\eta)\right\}^{\prime}.\label{cosgrove:DV_dif_eq2}
\end{align}
\end{theorem}
Comparison of the Hankelians formed by $ \beta_m(\eta)$ and the one in \eqref{cosgrove:chi_case_2} shows  that both are identical, provided $p \rightarrow 0$.\footnote{Note, however, that in this case, the expression in the curly brackets in~\eqref{cosgrove:chi_case_2} is clearly singular.} 

The differential equation from Theorem~\ref{cosgrove:Dale_theorem2} was investigated using other methods in \cite{Chalkley:1994rg}, and it was found that the persymmetric determinant $\hat{\mathcal{H}}_{n}=\hat{\mathcal{H}}_{n}=\vert {\beta}_m(x,y;p) \vert_{n};\quad m = 0, 1, \dots, 2n - 2$ can be represented by a finite polynomial.\footnote{More precisely, the solution is a finite-degree polynomial of degree \( 2n - 2 \):
\[
y(x) = K(n) \sum_{k=0}^{2n-2} p(n,k) x^k,
\]
where:
\begin{align}
   K(n) = \frac{n}{(n-1)!} \prod_{j=1}^{n-1} \frac{j!^3}{(n + j)!}; \quad
  p(n,k) = \sum_{r=0}^k \frac{(2r + 1 - k)}{(r + 1)\left[r!\right]^{2}\left[(k - r)!\right]^{2}} \prod_{j=1}^r (n^2 - j^2) \prod_{s=0}^{k - r - 1} (n^2 - s^2).
\end{align}
In the process of verifying this solution, we found that it can also be expressed in terms of the Appell \( \mathrm{F}_{4} \) function
\cite{Erdelyi:1981dy,Slater:2008wf}:
\begin{align}
    y(x)=K(n) \cdot \rm{F_4}\left( -n + 1, n + 1; 1, 1; x, -x \right).
\end{align}
} Both differential equations share a strikingly similar structure, and the explicit closed-form polynomial solution to the Cosgrove-Dale equation remains an open problem, particularly significant in the context of Nakamura’s conjecture. Both equations naturally arise from our general Hankelian through the procedure outlined above, and uncovering the precise relationship between them presents another intriguing open problem.

\section{Conclusion}
\label{con}
To summarize, we have found a solution to the Ernst equation that naturally includes both the Yamazaki–Hori and Vein solutions as special cases. The family of solutions is parametrized by an arbitrary function \( h \), which can be chosen to simplify the analysis. For a specific choice of \( h \), it is possible to establish connections with known equations, such as the Cosgrove–Dale equation \eqref{intro:Cosgrove_H4}.

The results presented in this paper will serve as the foundation for a forthcoming publication aimed at investigating Nakamura’s conjecture on the equivalence between the Ernst equation and the two-dimensional Toda model. A possible connection between \( h(x, y) \) in our solution and \( \psi(x, y) \) in the two-dimensional Toda model remains to be established and presents an interesting problem for further investigation.



\bibliographystyle{elsarticle-num}
\bibliography{ernst_eq}

\begin{thebibliography}{10}
\expandafter\ifx\csname url\endcsname\relax
  \def\url#1{\texttt{#1}}\fi
\expandafter\ifx\csname urlprefix\endcsname\relax\def\urlprefix{URL }\fi
\expandafter\ifx\csname href\endcsname\relax
  \def\href#1#2{#2} \def\path#1{#1}\fi

\bibitem{Weyl:1917tb}
H.~Weyl, {Zur gravitationstheorie}, Ann. Phys. 359~(18) (1917) 117--145.
\newblock \href {https://doi.org/10.1002/andp.19173591804} {\path{doi:10.1002/andp.19173591804}}.

\bibitem{Lewis:1932ln}
T.~Lewis, {Some special solutions of the equations of axially symmetric gravitational fields}, Proc. R. Soc. Lond. A Math. Phys. Sci. 136~(829) (1932) 176--192.
\newblock \href {https://doi.org/10.1098/rspa.1932.0073} {\path{doi:10.1098/rspa.1932.0073}}.

\bibitem{Papapetrou:1953db}
A.~Papapetrou, {Eine rotationssymmetrische Lösung in der allgemeinen Relativitätstheorie}, Annalen der Physik 447 (1953) 309--315.
\newblock \href {https://doi.org/10.1002/ANDP.19534470412} {\path{doi:10.1002/ANDP.19534470412}}.

\bibitem{Islam:2009no}
J.~N. Islam, {Rotating fields in general relativity}, Cambridge University Press, Cambridge, England, 2009.
\newblock \href {https://doi.org/10.1017/cbo9780511735738} {\path{doi:10.1017/cbo9780511735738}}.

\bibitem{Batic:2023ye}
D.~Batic, {The Tomimatsu–Sato metric reloaded}, Universe 9~(2) (2023) 77.
\newblock \href {https://doi.org/10.3390/universe9020077} {\path{doi:10.3390/universe9020077}}.

\bibitem{Tomimatsu:1973sk}
A.~Tomimatsu, H.~Sato, {New series of exact solutions for gravitational fields of spinning masses}, Prog. Theor. Phys. 50~(1) (1973) 95--110.
\newblock \href {https://doi.org/10.1143/PTP.50.95} {\path{doi:10.1143/PTP.50.95}}.

\bibitem{Tomimatsu:1972tg}
A.~Tomimatsu, H.~Sato, {New exact solution for the gravitational field of a spinning mass}, Phys. Rev. Lett. 29~(19) (1972) 1344--1345.
\newblock \href {https://doi.org/10.1103/physrevlett.29.1344} {\path{doi:10.1103/physrevlett.29.1344}}.

\bibitem{Kodama:2003en}
H.~Kodama, W.~Hikida, {Global structure of the Zipoy–Voorhees–Weyl spacetime and the = 2 Tomimatsu–Sato spacetime}, Class. Quantum Gravity 20~(23) (2003) 5121--5140.
\newblock \href {https://doi.org/10.1088/0264-9381/20/23/011} {\path{doi:10.1088/0264-9381/20/23/011}}.

\bibitem{Hoenselaers:1979vq}
C.~Hoenselaers, {Weyl conform tensor of the Tomimatsu-Sato $\delta = 3$ metric}, Gen. Relativ. Gravit. 11~(5) (1979) 325--327.
\newblock \href {https://doi.org/10.1007/bf00759273} {\path{doi:10.1007/bf00759273}}.

\bibitem{Gibbons:1973zo}
G.~W. Gibbons, R.~A. Russell-Clark, {Note on the Sato-Tomimatsu solution of Einstein's equations}, Phys. Rev. Lett. 30~(9) (1973) 398--399.
\newblock \href {https://doi.org/10.1103/physrevlett.30.398} {\path{doi:10.1103/physrevlett.30.398}}.

\bibitem{Manko:2012jv}
V.~S. Manko, {On the physical interpretation of $\delta= 2$ tomimatsu-Sato solution}, Prog. Theor. Phys. 127~(6) (2012) 1057--1075.
\newblock \href {https://doi.org/10.1143/ptp.127.1057} {\path{doi:10.1143/ptp.127.1057}}.

\bibitem{Tanabe:1976gp}
Y.~Tanabe, {Multipole moments in general relativity}, Prog Theor Phys 55~(1) (1976) 106--114.
\newblock \href {https://doi.org/10.1143/PTP.55.106} {\path{doi:10.1143/PTP.55.106}}.

\bibitem{Tanabe:1976mc}
Y.~Tanabe, {Multipole moments of the Tomimatsu-Sato gravitational fields}, Prog. Theor. Phys. 55~(3) (1976) 980--981.
\newblock \href {https://doi.org/10.1143/ptp.55.980} {\path{doi:10.1143/ptp.55.980}}.

\bibitem{Yamazaki:1977dp}
M.~Yamazaki, {On the Kerr–Tomimatsu–Sato family of spinning mass solutions}, J. Math. Phys. 18~(12) (1977) 2502--2508.
\newblock \href {https://doi.org/10.1063/1.523213} {\path{doi:10.1063/1.523213}}.

\bibitem{Yamazaki:1977xd}
M.~Yamazaki, S.~Hori, {Generalization of the Tomimatsu-Sato solutions}, Prog. Theor. Phys. 57~(2) (1977) 696--697.
\newblock \href {https://doi.org/10.1143/ptp.57.696} {\path{doi:10.1143/ptp.57.696}}.

\bibitem{Hori:1978ie}
S.~Hori, {On the exact solution of Tomimatsu-Sato family for an arbitrary integral value of the deformation parameter}, Prog. Theor. Phys. 59~(6) (1978) 1870--1891.
\newblock \href {https://doi.org/10.1143/ptp.59.1870} {\path{doi:10.1143/ptp.59.1870}}.

\bibitem{Hori:1996iz}
S.~Hori, {Generalization of Tomimatsu-Sato Solutions. I}, Prog. Theor. Phys. 95~(1) (1996) 65--70.
\newblock \href {https://doi.org/10.1143/ptp.95.65} {\path{doi:10.1143/ptp.95.65}}.

\bibitem{Hori:1996bb}
S.~Hori, {Generalization of tomimatsu-Sato solutions. {II}}, Prog. Theor. Phys. 95~(3) (1996) 557--564.
\newblock \href {https://doi.org/10.1143/ptp.95.557} {\path{doi:10.1143/ptp.95.557}}.

\bibitem{Hori:1996vt}
S.~Hori, {Generalization of tomimatsu-Sato solutions. {III}}, Prog. Theor. Phys. 95~(6) (1996) 1097--1120.
\newblock \href {https://doi.org/10.1143/ptp.95.1097} {\path{doi:10.1143/ptp.95.1097}}.

\bibitem{Hori:1996kp}
S.~Hori, {Generalization of tomimatsu-Sato solutions. {IV}}, Prog. Theor. Phys. 96~(2) (1996) 327--345.
\newblock \href {https://doi.org/10.1143/ptp.96.327} {\path{doi:10.1143/ptp.96.327}}.

\bibitem{Cosgrove:1978jo}
C.~Cosgrove, {A new formulation of the field equations for the stationary axisymmetric vacuum gravitational field. I. General theory}, Journal of Physics A 11 (1978) 2389--2404.
\newblock \href {https://doi.org/10.1088/0305-4470/11/12/007} {\path{doi:10.1088/0305-4470/11/12/007}}.

\bibitem{Cosgrove:1978cp}
C.~M. Cosgrove, {A new formulation of the field equations for the stationary axisymmetric vacuum gravitational field. II. Separable solutions}, J. Phys. A Math. Gen. 11~(12) (1978) 2405--2430.
\newblock \href {https://doi.org/10.1088/0305-4470/11/12/008} {\path{doi:10.1088/0305-4470/11/12/008}}.

\bibitem{Cosgrove:2008rh}
C.~M. Cosgrove, {Stationary axisymmetric gravitational fields: An asymptotic flatness preserving transformation}, in: {Lecture Notes in Physics}, Springer Berlin Heidelberg, Berlin, Heidelberg, 2008, pp. 444--453.
\newblock \href {https://doi.org/10.1007/3-540-09992-1\_116} {\path{doi:10.1007/3-540-09992-1\_116}}.

\bibitem{Cosgrove:1977ul}
C.~M. Cosgrove, {Limits of the generalised Tomimatsu-Sato gravitational fields}, J. Phys. A Math. Gen. 10~(12) (1977).

\bibitem{Cosgrove:1981zi}
C.~M. Cosgrove, {Bäcklund transformations in the Hauser–Ernst formalism for stationary axisymmetric spacetimes}, J. Math. Phys. 22~(11) (1981) 2624--2639.
\newblock \href {https://doi.org/10.1063/1.524841} {\path{doi:10.1063/1.524841}}.

\bibitem{Cosgrove:1982ol}
C.~M. Cosgrove, {Relationship between the inverse scattering techniques of Belinskii–Zakharov and Hauser–Ernst in general relativity}, J. Math. Phys. 23~(4) (1982) 615--633.
\newblock \href {https://doi.org/10.1063/1.525399} {\path{doi:10.1063/1.525399}}.

\bibitem{Cosgrove:1977np}
C.~M. Cosgrove, {New family of exact stationary axisymmetric gravitational fields generalising the Tomimatsu-Sato solutions}, J. Phys. A Math. Gen. 10~(9) (1977) 1481--1524.
\newblock \href {https://doi.org/10.1088/0305-4470/10/9/010} {\path{doi:10.1088/0305-4470/10/9/010}}.

\bibitem{Cosgrove:1980wx}
C.~M. Cosgrove, {Relationships between the group-theoretic and soliton-theoretic techniques for generating stationary axisymmetric gravitational solutions}, J. Math. Phys. 21~(9) (1980) 2417--2447.
\newblock \href {https://doi.org/10.1063/1.524680} {\path{doi:10.1063/1.524680}}.

\bibitem{Dale:1978xx}
P.~Dale, {Axisymmetric gravitational fields: a nonlinear differential equation that admits a series of exact eigenfunction solutions}, Proc. R. Soc. Lond. 362~(1711) (1978) 463--468.
\newblock \href {https://doi.org/10.1098/rspa.1978.0144} {\path{doi:10.1098/rspa.1978.0144}}.

\bibitem{Nakamura:1993ig}
A.~Nakamura, {Relation between tomimatsu-Sato black holes and semi-infinite Toda lattices}, J. Phys. Soc. Jpn. 62~(1) (1993) 368--369.
\newblock \href {https://doi.org/10.1143/jpsj.62.368} {\path{doi:10.1143/jpsj.62.368}}.

\bibitem{Nakamura:1991sv}
A.~Nakamura, Y.~Ohta, {Bilinear, Pfaffian and Legendre function structures of the Tomimatsu-Sato solutions of the Ernst equation in general relativity}, J. Phys. Soc. Jpn. 60~(6) (1991) 1835--1838.
\newblock \href {https://doi.org/10.1143/jpsj.60.1835} {\path{doi:10.1143/jpsj.60.1835}}.

\bibitem{Fukuyama:1995cm}
T.~Fukuyama, K.~Kamimura, Y.~U. Songju, {Toda Lattice and Tomimatsu-Sato Solutions}, J. Phys. Soc. Jpn. 64~(9) (May 1995).
\newblock \href {https://doi.org/10.1143/JPSJ.64.3201} {\path{doi:10.1143/JPSJ.64.3201}}.

\bibitem{Fukuyama:2011ma}
T.~Fukuyama, K.~Koizumi, {Toda molecule and Tomimatsu–Sato solution—towards the complete proof of Nakamura's conjecture}, J. Phys. A Math. Theor. 44~(34) (2011) 345201.
\newblock \href {https://doi.org/10.1088/1751-8113/44/34/345201} {\path{doi:10.1088/1751-8113/44/34/345201}}.

\bibitem{Hirota:1988wz}
R.~Hirota, {Toda Molecule Equations}, in: {Algebraic Analysis}, Elsevier, 1988, pp. 203--216.
\newblock \href {https://doi.org/10.1016/b978-0-12-400465-8.50024-9} {\path{doi:10.1016/b978-0-12-400465-8.50024-9}}.

\bibitem{Hirota:1988tt}
R.~Hirota, M.~Ito, F.~Kako, {Two-Dimensional Toda Lattice Equations}, Progr. Theoret. Phys. Suppl. 94 (1988) 42--58.
\newblock \href {https://doi.org/10.1143/ptps.94.42} {\path{doi:10.1143/ptps.94.42}}.

\bibitem{Ueno:2018gj}
K.~Ueno, K.~Takasaki, {Toda Lattice Hierarchy}, in: {Group Representations and Systems of Differential Equations}, Vol.~4, Mathematical Society of Japan, Tokyo, Japan, 2018, pp. 1--96.
\newblock \href {https://doi.org/10.2969/aspm/00410001} {\path{doi:10.2969/aspm/00410001}}.

\bibitem{Hirota:1988de}
R.~Hirota, Y.~Ohta, J.~Satsuma, {Wronskian Structures of Solutions for Soliton Equations}, Progr. Theoret. Phys. Suppl. 94 (1988) 59--72.
\newblock \href {https://doi.org/10.1143/ptps.94.59} {\path{doi:10.1143/ptps.94.59}}.

\bibitem{Hirota:2009fa}
R.~Hirota, {The direct method in soliton theory}, Cambridge University Press, Cambridge, England, 2009.
\newblock \href {https://doi.org/10.1017/cbo9780511543043} {\path{doi:10.1017/cbo9780511543043}}.

\bibitem{Vein:1983if}
P.~R. Vein, {Persymmetric determinants 4. An alternative form of the Yamazaki-Hori determinantal solution of the Ernst equation}, Linear Multilinear Algebra 12~(4) (1983) 329--339.
\newblock \href {https://doi.org/10.1080/03081088308817498} {\path{doi:10.1080/03081088308817498}}.

\bibitem{Vein:1980qb}
P.~R. Vein, P.~Dale, {Determinants, their derivatives and nonlinear differential equations}, J. Math. Anal. Appl. 74~(2) (1980) 599--634.
\newblock \href {https://doi.org/10.1016/0022-247x(80)90150-x} {\path{doi:10.1016/0022-247x(80)90150-x}}.

\bibitem{Vein:1982jk}
P.~R. Vein, {A short survey of some recent applications of determinants}, Linear Algebra Appl. 42 (1982) 287--297.
\newblock \href {https://doi.org/10.1016/0024-3795(82)90157-4} {\path{doi:10.1016/0024-3795(82)90157-4}}.

\bibitem{Vein:2019nj}
P.~R. Vein, {Persymmetric determinants}, Linear Multilinear Algebra 11~(3) (2019) 267--276.
\newblock \href {https://doi.org/10.1080/778390292} {\path{doi:10.1080/778390292}}.

\bibitem{Vein:1982ga}
P.~R. Vein, {Persymmetric determinants 3: A basic determinant}, Linear Multilinear Algebra 11~(4) (1982) 305--315.
\newblock \href {https://doi.org/10.1080/03081088208817456} {\path{doi:10.1080/03081088208817456}}.

\bibitem{Vein:1983ng}
P.~R. Vein, {Persymmetric determinants 5. Families of overlapping coaxial equivalent determinants}, Linear Multilinear Algebra 14~(2) (1983) 131--141.
\newblock \href {https://doi.org/10.1080/03081088308817550} {\path{doi:10.1080/03081088308817550}}.

\bibitem{Vein:1982xs}
P.~R. Vein, {Persymmetric determinants: The derivatives of determinants with Appell function elements}, Linear Multilinear Algebra 11~(3) (1982) 253--265.
\newblock \href {https://doi.org/10.1080/03081088208817448} {\path{doi:10.1080/03081088208817448}}.

\bibitem{Vein:2013kv}
R.~Vein, P.~Dale, {Determinants and their applications in mathematical physics}, Applied Mathematical Sciences, Springer, New York, NY, 2013.

\bibitem{Muir:2023zs}
T.~Muir, {A treatise on the theory of determinants}, Legare Street Press, 2023.

\bibitem{Zhang:2005cb}
F.~Zhang (Ed.), {The Schur complement and its applications}, 2005th Edition, Numerical Methods and Algorithms, Springer, New York, NY, 2005.
\newblock \href {https://doi.org/10.1007/b105056} {\path{doi:10.1007/b105056}}.

\bibitem{Call:1993vh}
G.~S. Call, D.~J. Velleman, {Pascal's Matrices}, Am. Math. Mon. 100~(4) (1993) 372--376.
\newblock \href {https://doi.org/10.1080/00029890.1993.11990415} {\path{doi:10.1080/00029890.1993.11990415}}.

\bibitem{Zhang:1998ub}
Z.~Zhang, M.~Liu, {An extension of the generalized Pascal matrix and its algebraic properties}, Linear Algebra Appl. 271~(1-3) (1998) 169--177.
\newblock \href {https://doi.org/10.1016/s0024-3795(97)00266-8} {\path{doi:10.1016/s0024-3795(97)00266-8}}.

\bibitem{Zhang:1997ah}
Z.~Zhang, {The linear algebra of the generalized Pascal matrix}, Linear Algebra Appl. 250 (1997) 51--60.
\newblock \href {https://doi.org/10.1016/0024-3795(95)00452-1} {\path{doi:10.1016/0024-3795(95)00452-1}}.

\bibitem{Stanimirovic:2011py}
S.~Stanimirović, {A generalization of the Pascal matrix and its properties}, Facta Univ. Ser. Math. Inform. 26 (2011) 17--27.

\bibitem{Harnad:2021av}
J.~Harnad, F.~Balogh, {Tau functions and their applications}, Cambridge Monographs on Mathematical Physics, Cambridge University Press, Cambridge, England, 2021.
\newblock \href {https://doi.org/10.1017/9781108610902} {\path{doi:10.1017/9781108610902}}.

\bibitem{Leznov:1981bf}
A.~N. Leznov, M.~V. Saveliev, {Theory of group representations and integration of nonlinear systems xa, zz=exp(kx)a}, Physica D 3~(1-2) (1981) 62--72.
\newblock \href {https://doi.org/10.1016/0167-2789(81)90119-6} {\path{doi:10.1016/0167-2789(81)90119-6}}.

\bibitem{Okamoto:1986cf}
K.~Okamoto, {Studies on the Painlevé equations: I.-Sixth Painlevé equation {PVI}}, Ann. Mat. Pura Appl. (4) 146~(1) (1986) 337--381.
\newblock \href {https://doi.org/10.1007/bf01762370} {\path{doi:10.1007/bf01762370}}.

\bibitem{Burchnall:1954ug}
J.~L. Burchnall, {A method of evaluating certain determinants}, Proc. Edinb. Math. Soc. (2) 9~(2) (1954) 100--104.
\newblock \href {https://doi.org/10.1017/s0013091500021362} {\path{doi:10.1017/s0013091500021362}}.

\bibitem{Chalkley:1994rg}
R.~Chalkley, {A persymmetric determinant}, J. Math. Anal. Appl. 187~(1) (1994) 107--117.
\newblock \href {https://doi.org/10.1006/jmaa.1994.1347} {\path{doi:10.1006/jmaa.1994.1347}}.

\bibitem{Erdelyi:1981dy}
A.~Erdelyi, {Higher Transcendental Functions}, Krieger Publishing Company, Melbourne, FL, 1981.

\bibitem{Slater:2008wf}
L.~J. Slater, {Generalized Hypergeometric Functions}, Cambridge University Press, Cambridge, England, 2008.

\end{thebibliography}





\end{document}